\documentclass{ws-procs961x669}            % default, citations in superscript
\begin{document}
\title{Dark Energy Density in SUGRA models and degenerate vacua}

\author{C. D. Froggatt}

\address{School of Physics and Astronomy, University of Glasgow,\\
Glasgow, G12 8QQ, UK\\
E-mail: Colin.Froggatt@glasgow.ac.uk
}

\author{H. B. Nielsen}

\address{The Niels Bohr Institute, University of Copenhagen, Blegdamsvej 17\\
Copenhagen, DK-2100, Denmark\\
E-mail: hbech@nbi.dk
}

\author{R. Nevzorov$^*$ and A. W. Thomas}

\address{ARC Centre of Excellence for Particle Physics at the Terascale and CSSM,\\
School of Physical Sciences, The University of Adelaide,\\
Adelaide, SA 5005, Australia\\
$^*$E-mail: roman.nevzorov@adelaide.edu.au}

\begin{abstract}
In $N=1$ supergravity the tree-level scalar potential of the hidden sector may have
a minimum with broken local supersymmetry (SUSY) as well as a supersymmetric Minkowski vacuum.
These vacua can be degenerate, allowing for a consistent implementation of the
multiple point principle. The first minimum where SUSY is broken can be
identified with the physical phase in which we live. In the second supersymmetric phase,
in flat Minkowski space, SUSY may be broken dynamically either in the observable or in
the hidden sectors inducing a tiny vacuum energy density. We argue that the exact
degeneracy of these phases may shed light on the smallness of the cosmological
constant. Other possible phenomenological implications are also discussed.
In particular, we point out that the presence of such degenerate vacua may lead to
small values of the quartic Higgs coupling and its beta function at the Planck scale
in the physical phase.
\end{abstract}


\bodymatter

\section{Introduction}
It is expected that at ultra-high energies the Standard Model (SM) is
embedded in an underlying theory that provides a framework for the unification
of all interactions such as Grand Unified Theories (GUTs), supergravity (SUGRA),
String Theory, etc. At low energies this underlying theory could lead to new physics
phenomena beyond the SM. Moreover the energy scale associated with the physics
beyond the SM is supposed to be somewhat close to the mass of the Higgs boson
to avoid a fine-tuning problem related to the need to stabilize the scale where
electroweak (EW) symmetry is broken.
Despite the successful discovery of the $125\,\mbox{GeV}$ Higgs boson in 2012,
no indication of any physics beyond the SM has been detected at the LHC so far.
On the other hand, there are compelling reasons to believe that the SM is extremely
fine-tuned. Indeed, astrophysical and cosmological observations indicate that there
is a tiny energy density spread all over the Universe (the cosmological constant), i.e.
$\rho_{\Lambda} \sim 10^{-123}M_{Pl}^4 \sim 10^{-55} M_Z^4$ \cite{Bennett:2003bz, Spergel:2003cb},
which is responsible for its acceleration. At the same time much bigger contributions
must come from the EW symmetry breaking ($\sim 10^{-67}M_{Pl}^4$) and
QCD condensates ($\sim 10^{-79}M_{Pl}^4$).
Because of the enormous cancellation between the contributions of different condensates to
$\rho_{\Lambda}$, which is required to keep $\rho_{\Lambda}$ around its measured value,
the smallness of the cosmological constant can be considered as a fine-tuning problem.

Here, instead of trying to alleviate fine-tuning of the SM we impose the exact degeneracy of
at least two (or even more) vacua. Their presence was predicted by the so-called Multiple Point
Principle (MPP) \cite{Bennett:1993pj}. According to the MPP, Nature chooses values
of coupling constants such that many phases of the underlying theory should coexist. This
corresponds to a special (multiple) point on the phase diagram where these phases meet.
At the multiple point these different phases have the same vacuum energy density.

The MPP applied to the SM implies that the Higgs effective potential, which can be written as
\begin{equation}
V_{eff}(H) = m^2(\phi) H^{\dagger} H + \lambda(\phi) (H^{\dagger} H)^2\,,
\label{11}
\end{equation}
where $H$ is a Higgs doublet and $\phi$ is a norm of the Higgs field, i.e. $\phi^2=H^{\dagger} H$,
has two degenerate minima. These minima are taken to be at the EW and Planck scales \cite{Froggatt:1995rt}.
The corresponding vacua can have the same energy density only if
\begin{equation}
\lambda(M_{Pl})\simeq 0 \,, \qquad\quad \beta_{\lambda}(M_{Pl})\simeq 0\,.
\label{12}
\end{equation}
where $\beta_{\lambda}=\frac{d \lambda(\phi)}{d \log\phi}$ is the beta--function of
$\lambda(\phi)$. It was shown that the MPP conditions (\ref{12}) can be satisfied when $M_t=173\pm 5\,\mbox{GeV}$ and
$M_H=135\pm 9\, \mbox{GeV}$ \cite{Froggatt:1995rt}. The application of the MPP to the two Higgs doublet extension of
the SM was also considered \cite{Froggatt:2006zc, Froggatt:2007qp, Froggatt:2008am}.

The measurement of the Higgs boson mass allows us to determine quite precisely the parameters of the
Higgs potential (\ref{11}). Furthermore using the extrapolation of the SM parameters up to $M_{Pl}$  with
full 3--loop precision it was found \cite{Buttazzo:2013uya} that
\begin{equation}
\lambda(M_{Pl}) =  -0.0143-0.0066\left(\frac{M_t}{\mbox{GeV}} - 173.34 \right)\qquad\qquad\qquad\qquad
\label{13}
\end{equation}
\[
\qquad\qquad\qquad\qquad+ 0.0018\left(\frac{\alpha_3(M_Z)-0.1184}{0.0007}\right) +0.0029\left(\frac{M_H}{\mbox{GeV}} -125.15 \right)\,.
\]
The computed value of  $\beta_{\lambda}(M_{Pl})$ is also rather small, so that the MPP
conditions (\ref{12}) are basically fulfilled.

The successful MPP predictions for the Higgs and top quarks masses \cite{Froggatt:1995rt} suggest that
we may use this idea to explain the tiny value of the cosmological constant as well. In principle, the smallness
of the cosmological constant could be related to an almost exact symmetry. Nevertheless, none of the
generalizations of the SM provides any satisfactory explanation for the smallness of this dark energy density.
An exact global supersymmetry (SUSY) guarantees that the vacuum energy density vanishes in all global minima
of the scalar potential. However the non-observation of superpartners of quarks and leptons implies that
SUSY is broken. The breakdown of SUSY induces a huge and positive contribution to the dark energy density
which is many orders of magnitude larger than $M_Z^4$. Here the MPP assumption is adapted to $(N=1)$
SUGRA models, in order to provide an explanation for the tiny deviation of the measured dark energy density
from zero.

\section{SUGRA models inspired by degenerate vacua}
The full $(N=1)$ SUGRA Lagrangian can be specified
in terms of an analytic gauge kinetic function $f_a(\phi_{M})$ and
a real gauge-invariant K$\Ddot{a}$hler function
$G(\phi_{M},\phi_{M}^{*})$. These functions depend on the
chiral superfields $\phi_M$. The function $f_{a}(\phi_M)$
determine, in particular, the gauge couplings $Re f_a(\phi_M)=1/g_a^2$,
where the index $a$ represents different gauge groups.
The K$\Ddot{a}$hler function can be presented as
\begin{equation}
G(\phi_{M},\phi_{M}^{*})=K(\phi_{M},\phi_{M}^{*})+\ln|W(\phi_M)|^2\,,
\label{21}
\end{equation}
where $K(\phi_{M},\phi_{M}^{*})$ is the K$\Ddot{a}$hler potential
while $W(\phi_M)$ is the superpotential of the SUGRA model
under consideration. Here we shall use standard supergravity mass units:
$\frac{M_{Pl}}{\sqrt{8\pi}}=1$.

The SUGRA scalar potential can be written as a sum of $F-$ and $D$-terms, i.e.
\begin{equation}
\begin{array}{c}
V(\phi_M,\phi^{*}_M)=\sum_{M,\,\bar{N}} e^{G}\left(G_{M}G^{M\bar{N}}
G_{\bar{N}}-3\right)+ \frac{1}{2}\sum_{a}(D^{a})^2,\\[2mm]
D^{a}=g_{a}\sum_{i,\,j}\left(G_i T^a_{ij}\phi_j\right)\,,\qquad
G_M \equiv  \frac{\partial G}{\partial \phi_M}\,,\qquad
G_{\bar{M}}\equiv  \frac{\partial G}{\partial \phi^{*}_M}\,,\\[2mm]
G_{\bar{N}M}\equiv \frac{\partial^2 G}{\partial \phi^{*}_N \partial \phi_M}\,,\qquad\qquad
G^{M\bar{N}}=G_{\bar{N}M}^{-1}\,.
\end{array}
\label{22}
\end{equation}
In Eq.~(\ref{22}) $g_a$ is the gauge coupling associated with the
generator $T^a$. In order to achieve the breakdown of local SUSY
in $(N=1)$ supergravity, a hidden sector is introduced. The superfields of the hidden
sector $(z_i)$ interact with the observable ones only by means of
gravity. It is expected that at the minimum of the scalar potential (\ref{22}) hidden
sector fields acquire vacuum expectation values (VEVs) so that at least one of their auxiliary fields
\begin{equation}
F^{M}=e^{G/2}G^{M\bar{P}}G_{\bar{P}}
\label{23}
\end{equation}
is non-vanishing, leading to the spontaneous breakdown of local SUSY,
giving rise to a non-zero gravitino mass $m_{3/2}\simeq <e^{G/2}>$.
The absolute value of the vacuum energy density at the minimum of the SUGRA
scalar potential (\ref{22}) tends to be of order of $m_{3/2}^2 M_{Pl}^2$.
Therefore an enormous degree of fine-tuning is required to keep the cosmological
constant in SUGRA models around its observed value.

The successful implementation of the MPP in $(N=1)$ SUGRA models requires
us to assume the existence of a supersymmetric Minkowski vacuum\cite{Froggatt:2003jm, Froggatt:2005nb}.
According to the MPP this second vacuum and the physical one must be degenerate.
Since the vacuum energy density of supersymmetric states in flat Minkowski space vanishes,
the cosmological constant problem is solved to first approximation.
Such a second vacuum exists if the SUGRA scalar potential (\ref{22}) has
a minimum where
\begin{equation}
W(z_m^{(2)})=0\,,\qquad
\frac{\partial W(z_i)}{\partial z_m}\Biggl|_{z_m=z_m^{(2)}}=0
\label{24}
\end{equation}
where $z_m^{(2)}$ are VEVs of the hidden sector fields in the second vacuum.
Eqs.~(\ref{24}) indicate that an extra fine-tuning is needed to ensure the presence
of the supersymmetric Minkowski vacuum in SUGRA models.

The simplest K$\Ddot{a}$hler potential and superpotential
that satisfies conditions (\ref{24}) are given by
\begin{equation}
K(z,\,z^{*})=|z|^2\,,\qquad\qquad W(z)=m_0(z+\beta)^2\,.
\label{25}
\end{equation}
The hidden sector of the corresponding SUGRA model involves only one singlet
superfield $z$. If $\beta=\beta_0=-\sqrt{3}+2\sqrt{2}$, the corresponding
SUGRA scalar potential possesses two degenerate vacua with zero energy density
at the classical level. The first minimum associated with $z^{(2)}=-\beta$
is a supersymmetric Minkowski vacuum. In the other minimum, $z^{(1)}=\sqrt{3}-\sqrt{2}$,
local SUSY is broken so that it can be identified with the physical vacuum.
Varying $\beta$ around $\beta_0$ one can get a positive or a negative contribution
from the hidden sector to the total vacuum energy density of the physical phase.
Thus parameter $\beta$ can be always fine-tuned so that the
physical and supersymmetric Minkowski vacua are degenerate.

The fine-tuning associated with the realisation of the MPP in $(N=1)$ SUGRA
models can be to some extent alleviated within the no-scale inspired SUGRA model
with broken dilatation invariance \cite{Froggatt:2005nb}.
Let us consider the no--scale inspired SUGRA model that involves two hidden sector
superfields ($T$ and $z$) and a set of chiral supermultiplets $\varphi_{\sigma}$ in
the observable sector. These superfields transform differently under the
dilatations ($T\to\alpha^2 T,\, z\to\alpha\,z,\,\varphi_{\sigma}\to\alpha\,\varphi_{\sigma}$)
and imaginary translations ($T\to T+i\beta,\, z\to z,\, \varphi_{\sigma}\to \varphi_{\sigma}$),
which are subgroups of the $SU(1,1)$ group \cite{Froggatt:2005nb, Froggatt:2004gc}.
The full superpotential of the model can be written as a sum \cite{Froggatt:2005nb}:
\begin{equation}
\begin{array}{c}
W(z,\,\varphi_{\alpha})=W_{hid} + W_{obs}\,,\\[1mm]
W_{hid}=\varkappa\biggl(z^3+\mu_0 z^2+\sum_{n=4}^{\infty}c_n z^n\biggr),\qquad
W_{obs}=\sum_{\sigma,\beta,\gamma}\frac{1}{6}
Y_{\sigma\beta\gamma}\varphi_{\sigma}\varphi_{\beta}\varphi_{\gamma}\,.
\end{array}
\label{26}
\end{equation}
The superpotential (\ref{26}) includes a bilinear mass term for the
superfield $z$ and higher order terms $c_n z^n$ which explicitly break dilatation
invariance. A term proportional to $z$ is not included since it can be forbidden
by a gauge symmetry of the hidden sector. Here we do not allow the breakdown
of dilatation invariance in the superpotential of the observable sector to avoid
potentially dangerous terms that may lead to the so--called $\mu$--problem, etc.

The full K$\ddot{a}$hler potential of the SUGRA model under consideration is
given by \cite{Froggatt:2005nb}:
\begin{equation}
\begin{array}{rcl}
K(\phi_{M},\phi_{M}^{*})&=&-3\ln\biggl[T+\overline{T}
-|z|^2-\sum_{\alpha}\zeta_{\alpha}|\varphi_{\alpha}|^2\biggr]
+\\[2mm]
&+&\sum_{\alpha, \beta}\biggl(\frac{\eta_{\alpha\beta}}
{2}\,\varphi_{\alpha}\,\varphi_{\beta}+h.c.\biggr)+
\sum_{\beta}\xi_{\beta}|\varphi_{\beta}|^2\,,
\end{array}
\label{27}
\end{equation}
where $\zeta_{\alpha}$, $\eta_{\alpha\beta}$, $\xi_{\beta}$ are some
constants. If $\eta_{\alpha\beta}$ and $\xi_{\beta}$ have non-zero values
the dilatation invariance is explicitly broken in the K$\ddot{a}$hler
potential of the observable sector. Here we restrict our consideration to
the simplest set of terms that break dilatation invariance. Moreover we only
allow the breakdown of the dilatation invariance in the K$\ddot{a}$hler
potential of the observable sector, because any variations in the part of
the K$\ddot{a}$hler potential associated with the hidden sector may spoil
the vanishing of the vacuum energy density in global minima.
When the parameters $\eta_{\alpha\beta}$, $\xi_{\beta}$
and $\varkappa$ go to zero, the dilatation invariance is restored,
protecting supersymmetry and a zero value of the cosmological constant.

In the SUGRA model under consideration the tree-level scalar potential
of the hidden sector
is positive definite
\begin{equation}
V_{hid}=\frac{1}{3(T+\overline{T}-|z|^2)^2}
\biggl|\frac{\partial W_{hid}(z)}{\partial z}\biggr|^2,
\label{29}
\end{equation}
so that the vacuum energy density vanishes near its global minima.
In the simplest case when $c_n=0$, the SUGRA scalar potential of the
hidden sector (\ref{29}) has two minima, at $z=0$ and $z=-\frac{2\mu_0}{3}$.
At these points $V_{hid}$ attains its absolute minimal value i.e.~zero.
In the first vacuum, where $z=-\frac{2\mu_0}{3}$,
local SUSY is broken and the gravitino gains mass
\begin{equation}
m_{3/2}=\biggl<\frac{W_{hid}(z)}{(T+\overline{T}-|z|^2)^{3/2}}\biggr>
=\frac{4\varkappa\mu_0^3}{27\biggl<\biggl(T+\overline{T}
-\frac{4\mu_0^2}{9}\biggr)^{3/2}\biggr>}\,.
\label{34}
\end{equation}
All scalar particles get non--zero masses
$m_{\sigma}\sim \frac{m_{3/2} \xi_{\sigma} }{\zeta_{\sigma}}$ as well.
This minimum can be identified with the physical vacuum.
Assuming that $\xi_{\alpha}$, $\zeta_{\alpha}$, $\mu_0$ and $<T>$ are all of
order unity, a SUSY breaking scale $M_S\sim 1\,\mbox{TeV}$
can only be obtained when $\varkappa$ is extremely small, i.e. $\varkappa\simeq 10^{-15}$.
In the second vacuum, where $z=0$, the superpotential of the hidden sector
vanishes and local SUSY remains intact giving rise to the supersymmetric Minkowski vacuum.
If the high order terms $c_n z^n$ are present in Eqs.~(\ref{26}),
$V_{hid}$ can have many degenerate vacua, with broken and unbroken SUSY,
in which the vacuum energy density vanishes. As a result the MPP conditions are fulfilled
without any extra fine-tuning at the tree--level.

It is worth noting that the inclusion of perturbative and non-perturbative corrections to the
Lagrangian of the no--scale inspired SUGRA model should spoil the degeneracy of vacua,
giving rise to a huge energy density in the minimum of the scalar potential where local SUSY is broken.
Furthermore, in the SUGRA model under consideration the mechanism for the stabilization of the
VEV of the hidden sector field $T$ remains unclear. Therefore this model should be considered
as a toy example only. It demonstrates that in $(N=1)$ supergravity there might be a mechanism
which ensures the vanishing of the vacuum energy density in the physical vacuum. This mechanism
can also result in a set of degenerate vacua with broken and unbroken SUSY,
leading to the realization of the MPP.

\section{Implications for cosmology and collider phenomenology}

\subsection{Model with intermediate SUSY breaking scale}
Now let us assume that the MPP inspired SUGRA model of the type just
discussed is realised in Nature. In other words there exist at least two
exactly degenerate phases. The first phase is associated with the physical
vacuum whereas the second one is identified with the supersymmetric
Minkowski vacuum in which the vacuum energy density vanishes in the leading
approximation. However non-perturbative effects may give rise to the
breakdown of SUSY in the second phase resulting in a small vacuum energy
density. This small energy density should be then transferred to our vacuum
by the assumed degeneracy.

If SUSY breaking takes place in the second vacuum, it can be caused by the
strong interactions in the observable sector. Indeed, the SM gauge couplings
$g_1$, $g_2$ and $g_3$, which correspond to $U(1)_Y$, $SU(2)_W$ and $SU(3)_C$
gauge interactions respectively, change with the energy scale. Their evolution
obeys the renormalization group equations (RGEs) that in the one--loop
approximation can be written as
\begin{equation}
\frac{d\log{\alpha_i(Q)}}{d\log{Q^2}}=\frac{b_i\alpha_i(Q)}{4\pi}\,,
\label{41}
\end{equation}
where $Q$ is a renormalization scale, $i=1,2,3$ and $\alpha_i(Q)=g_i^2(Q)/(4\pi)$.
In the pure MSSM $b_3<0$ and the gauge coupling $g_3(Q)$ of the $SU(3)_C$
interactions increases in the infrared region. Thus although this coupling can be
rather small at high energies it becomes rather large at low energies and
one can expect that the role of non--perturbative effects is enhanced.

To simplify our analysis we assume that the SM gauge couplings at high energies are
identical in the physical and supersymmetric Minkowski vacua. Consequently
for $Q>M_S$, where $M_S$ is a SUSY breaking scale in the physical vacuum,
the renormalization group (RG) flow of these couplings is also the same in both vacua.
When $Q<M_S$ all superparticles in the physical vacuum decouple. Therefore
the $SU(3)_C$ beta function in the physical and supersymmetric Minkowski vacua
become very different. Because of this, below the scale $M_S$ the values of $\alpha_3(Q)$
in the physical and second vacua ($\alpha^{(1)}_3(Q)$ and $\alpha^{(2)}_3(Q)$)
are not the same. For $Q<M_S$ the $SU(3)_C$ beta function in the physical
phase $\tilde{b}_3$ coincides with the corresponding SM beta function, i.e. $\tilde{b}_3=-7$.
Using the value of $\alpha^{(1)}_3(M_Z)\approx 0.1184$ and the matching condition
$\alpha^{(2)}_3(M_S)=\alpha^{(1)}_3(M_S)$, one can estimate the value of the
strong gauge coupling in the second vacuum
\begin{equation}
\frac{1}{\alpha^{(2)}_3(M_S)}=\frac{1}{\alpha^{(1)}_3(M_Z)}-
\frac{\tilde{b}_3}{4\pi}\ln\frac{M^2_{S}}{M_Z^2}\, .
\label{42}
\end{equation}

In the supersymmetric Minkowski vacuum all particles of the MSSM are massless and
the EW symmetry is unbroken. So, in the second phase the $SU(3)_C$ beta function $b_3$
remains the same as in the MSSM, i.e. $b_3=-3$. Since the MSSM $SU(3)_C$ beta function exhibits
asymptotically free behaviour, $\alpha^{(2)}_3(Q)$ increases in the infrared region.
The top quark is massless in the supersymmetric phase and its Yukawa coupling also grows in
the infrared with the increasing of $\alpha^{(2)}_3(Q)$. At the scale
\begin{equation}
\Lambda_{SQCD}=M_{S}\exp\left[{\frac{2\pi}{b_3\alpha_3^{(2)}(M_{S})}}\right]\,,
\label{43}
\end{equation}
where the supersymmetric QCD interactions become strong in the supersymmetric Minkowski
vacuum, the top quark Yukawa coupling is of the same order of magnitude as the $SU(3)_C$
gauge coupling. So a large value of the top quark Yukawa coupling may give rise to the formation
of a quark condensate. This condensate breaks SUSY, resulting in a non--zero positive value
for the dark energy density
\begin{equation}
\rho_{\Lambda} \simeq \Lambda_{SQCD}^4\, .
\label{44}
\end{equation}

\begin{figure}[h]
\begin{center}
\includegraphics[width=3.1in]{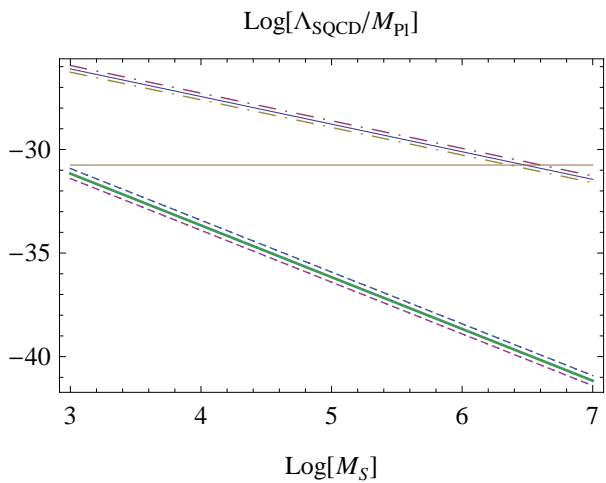}\\
\end{center}
\caption{The value of $\log\left[\Lambda_{SQCD}/M_{Pl}\right]$ versus
$\log M_S$. The thin and thick solid lines correspond to the pure MSSM and the
MSSM with an additional pair of $5+\bar{5}$ supermultiplets,
respectively. These lines are obtained for $\alpha_3(M_Z)=0.1184$.
The dashed and dash--dotted lines represent the uncertainty in $\alpha_3(M_Z)$.
The upper dashed and dash-dotted lines are obtained for $\alpha_3(M_Z)=0.121$,
while the lower ones correspond to $\alpha_3(M_Z)=0.116$. The horizontal line
represents the observed value of $\rho_{\Lambda}^{1/4}$. The SUSY
breaking scale, $M_S$, is measured in GeV.}
\label{fig1}
\end{figure}

The dependence of $\Lambda_{SQCD}$ on the SUSY breaking
scale $M_S$ is shown in Fig.~\ref{fig1}. Since $\tilde{b}_3 < b_3$ the value of
the QCD gauge coupling below $M_S$ is larger in the physical phase than in the
second one. Therefore the scale $\Lambda_{SQCD}$ is substantially lower
than the QCD scale in the SM and decreases with increasing $M_S$.
When the supersymmetry breaking scale in the physical phase is of the order
of $1\,\mbox{TeV}$, we get $\Lambda_{SQCD}=10^{-26}M_{Pl} \simeq 100$ eV
\cite{Froggatt:2005nb, Froggatt:2007qs, Froggatt:2009pj}.
This leads to the value of the cosmological constant
$\rho_{\Lambda} \simeq 10^{-104} M_{Pl}^4$, which is enormously suppressed
as compared with $v^4 \simeq 10^{-67} M_{Pl}^4$.  The measured value of the
dark energy density is reproduced when
$\Lambda_{SQCD}=10^{-31}M_{Pl} \simeq 10^{-3}\,\mbox{eV}$.
The appropriate values of $\Lambda_{SQCD}$ can be obtained only if
$M_S\simeq 10^3-10^4\,\mbox{TeV}$\cite{Froggatt:2005nb, Froggatt:2007qs, Froggatt:2009pj}.
However models with such a large SUSY breaking scale do not lead to
a suitable dark-matter candidate and also spoil the unification of the SM gauge couplings.

\subsection{Split SUSY scenario}
The problems mentioned above can be addressed within the Split SUSY scenario
of superymmetry breaking \cite{ArkaniHamed:2004fb, Giudice:2004tc}.
In other words, let us now assume that in the physical vacuum SUSY is broken so that
all scalar bosons gain masses of order of $M_S\gg 10\,\mbox{TeV}$, except for a SM-like
Higgs boson, whose mass is set to be around $125\,\mbox{GeV}$. The mass parameters
of gauginos and Higgsinos are protected by a combination of an R-symmetry and Peccei Quinn
symmetry so that they can be many orders of magnitude smaller than $M_S$.
To ensure gauge coupling unification all neutralino, chargino and gluino states
are chosen to lie near the TeV scale in the Split SUSY scenario \cite{Giudice:2004tc}.
Also a TeV-scale lightest neutralino can be an appropriate dark matter candidate
\cite{Giudice:2004tc, ArkaniHamed:2004yi, dm-split-susy-1, dm-split-susy-2}.

Thus in the Split SUSY scenario supersymmetry is not used to stabilize the EW scale
\cite{ArkaniHamed:2004fb, Giudice:2004tc}. This stabilization is expected to be
provided by some other mechanism, which may also explain the tiny value of
the dark energy density. Therefore in the Split SUSY models $M_S$ is taken to be much
above 10 TeV. In the Split SUSY scenario some flaws inherent to the MSSM
disappear. The ultra-heavy scalars, whose masses can range from hundreds
of TeV up to $10^{13}\,\mbox{GeV}$ \cite{Giudice:2004tc}, ensure the absence
of large flavor changing and CP violating effects. The stringent constraints from
flavour and electric dipole moment data, that require $M_S>100-1000\,\mbox{TeV}$,
are satisfied and the dimension-five operators, which mediate proton decay,
are also suppressed within the Split SUSY models.
Nevertheless, since the sfermions are ultra-heavy the Higgs sector is
extremely fine-tuned, with the understanding that the solution to both the
hierarchy and cosmological constant problems might not involve natural
cancellations, but follow from anthropic-like selection effects \cite{Weinberg:1987dv}.
In other words galaxy and star formation, chemistry and biology, are basically
impossible without these scales having the values found in our Universe
\cite{Weinberg:1987dv, anthropic-principle-1, anthropic-principle-2}. In this case
supersymmetry may be just a necessary ingredient in a fundamental theory of
Nature like in the case of String Theory.

It has been argued that String Theory can have a huge number of long-lived metastable vacua
\cite{landscape, landscape-1, landscape-2, Bousso:2000xa, large-number-of-vacua, large-number-of-vacua-1, large-number-of-vacua-2, large-number-of-vacua-3, high-energy-susy-breaking, high-energy-susy-breaking-1}
which is measured in googles ($\sim 10^{100}$)
\cite{Bousso:2000xa, large-number-of-vacua, large-number-of-vacua-1, large-number-of-vacua-2, large-number-of-vacua-3, high-energy-susy-breaking, high-energy-susy-breaking-1}.
The space of such vacua is called the ``landscape".  To analyze the huge multitude of universes,
associated with the ``landscape" of these vacua a statistical approach is used
\cite{large-number-of-vacua,  large-number-of-vacua-1, large-number-of-vacua-2, large-number-of-vacua-3, high-energy-susy-breaking, high-energy-susy-breaking-1}.
The total number of vacua in String Theory is sufficiently large to fine-tune both the cosmological
constant and the Higgs mass, favoring a high-scale breaking of supersymmetry \cite{high-energy-susy-breaking,  high-energy-susy-breaking-1}.
Thus it is possible for us to live in a universe fine-tuned in the way we find it simply because of a cosmic selection rule,
i.e. the anthropic principle \cite{Weinberg:1987dv}.

The idea of the multiple point principle
and the landscape paradigm have at least two things in common. Both approaches imply
the presence of a large number of vacua with broken and unbroken SUSY. The landscape
paradigm and MPP also imply that the parameters of the theory, which results in the SM
at low energies, can be extremely fine-tuned so as to guarantee a tiny vacuum energy
density and a large hierarchy between $M_{Pl}$ and the EW scale. Moreover the MPP
assumption might originate from the landscape of string theory vacua, if all vacua with
a vacuum energy density that is too large are forbidden for some reason, so that all the allowed
string vacua, with broken and unbroken supersymmetry, are degenerate to very high
accuracy. If this is the case, then the breaking of supersymmetry at high scales is perhaps
still favored. Although this scenario looks quite attractive it implies that only a narrow band
of values around zero cosmological constant would be allowed and the surviving vacua
would obey MPP to the accuracy of the width $w$ of this remaining band. However
such accuracy is not sufficient to become relevant for the main point of
the present article, according to which MPP ``transfers'' the vacuum energy density
of the second vacuum to the physical vacuum.

In order to estimate the value of the cosmological constant we again assume that
the physical and second phases have precisely the same vacuum energy densities
and the gauge couplings at high energies are identical in both vacua. This means
that the renormalization group flow of the SM gauge couplings down to the scale
$M_S$ is the same in both vacua as before. For $Q<M_S$ all squarks and sleptons
in the physical vacuum decouple and the beta functions change. At the TeV scale,
the corresponding beta functions in the physical phase change once again due to
the decoupling of the gluino, neutralino and chargino. Assuming that
$\alpha^{(2)}_3(M_S)=\alpha^{(1)}_3(M_S)$, one finds
\begin{equation}
\frac{1}{\alpha^{(2)}_3(M_S)}=\frac{1}{\alpha^{(1)}_3(M_Z)}-
\frac{\tilde{b}_3}{4\pi}\ln\frac{M^2_{g}}{M_Z^2}-\frac{b'_3}{4\pi}\ln\frac{M^2_{S}}{M_g^2}\,,
\label{45}
\end{equation}
where $M_g$ is the mass of the gluino and $b'_3=-5$ is the one--loop beta function of the strong
gauge coupling in the Split SUSY scenario. The values of $\Lambda_{SQCD}$ and $\rho_{\Lambda}$
can be estimated using Eqs.~(\ref{43}) and (\ref{44}), respectively.

\begin{figure}[h]
\begin{center}
\includegraphics[width=3.1in]{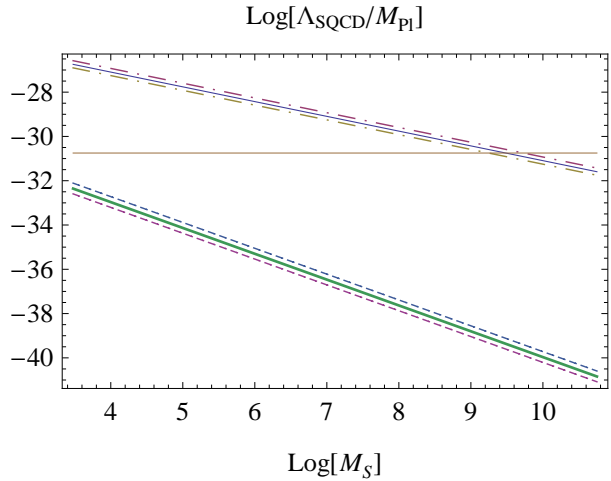}\\
\end{center}
\caption{The value of $\log\left[\Lambda_{SQCD}/M_{Pl}\right]$ versus $\log M_S$
for $M_q=M_g=3\,\mbox{TeV}$. The thin and thick solid lines correspond to the
Split SUSY scenarios with the pure MSSM particle content and the MSSM particle
content supplemented by an additional pair of $5+\bar{5}$ supermultiplets respectively.
The dashed and dash--dotted lines represent the uncertainty in $\alpha_3(M_Z)$.
The thin and thick solid lines are obtained for $\alpha_3(M_Z)=0.1184$,
the upper (lower) dashed and dash-dotted lines correspond to
$\alpha_3(M_Z)=0.116$ ($\alpha_3(M_Z)=0.121$). The horizontal line represents the
observed value of $\rho_{\Lambda}^{1/4}$. The SUSY breaking scale $M_S$ is measured in GeV.}
\label{fig2}
\end{figure}

In Fig.~\ref{fig2} we explore the dependence of $\Lambda_{SQCD}$ in the second phase
on the SUSY breaking scale $M_S$ in the physical vacuum. In our analysis we set $M_g=3\,\mbox{TeV}$.
As before $\Lambda_{SQCD}$ diminishes with increasing $M_S$. The observed value of the dark energy
density can be reproduced when $M_S\sim 10^9 - 10^{10}\,\mbox{GeV}$
\cite{Froggatt:2010iy, Froggatt:2011fc, Froggatt:2013aua}. The value of $M_S$, which
results in the measured cosmological constant, depends on $\alpha_3(M_Z)$ and the gluino mass.
However this dependence is rather weak. In particular, with increasing $M_g$ the value of $M_S$,
which leads to an appropriate value of the cosmological constant, decreases. When $\alpha_3(M_Z)=0.116-0.121$
and $M_g=500-2500\,\mbox{GeV}$, the corresponding value of the SUSY breaking scale varies from
$2\cdot 10^9\,\mbox{GeV}$ up to $3\cdot 10^{10}\,\mbox{GeV}$\cite{Froggatt:2010iy, Froggatt:2011fc, Froggatt:2013aua}.

The obtained prediction for $M_S$ can be tested. A striking feature of the Split SUSY model is the extremely
long lifetime of the gluino. The gluino decays through a virtual squark to a quark antiquark pair
and a neutralino $\tilde{g} \rightarrow q\bar{q}+\chi_1^0$. The large squark masses give rise to a long
lifetime for the gluino. This lifetime can be estimated as \cite{Dawson:1983fw, Hewett:2004nw}
\begin{equation}
\tau \sim 8\biggl(\frac{M_S}{10^9\,\mbox{GeV}}\biggr)^4
\biggl(\frac{1\,\mbox{TeV}}{M_g}\biggr)^5\,s.
\label{46}
\end{equation}
From Eq.~(\ref{46}) it follows that the supersymmetry breaking scale in the Split
SUSY models should not exceed $10^{13}\,\mbox{GeV}$ \cite{Giudice:2004tc}.
Otherwise the gluino lifetime becomes larger than the age of the Universe.
When $M_S$ varies from $2\cdot 10^9\,\mbox{GeV}$ ($M_g=2500\,\mbox{GeV}$)
to $3\cdot 10^{10}\,\mbox{GeV}$ ($M_g=500\,\mbox{GeV}$) the gluino lifetime
changes from $1\,\mbox{sec.}$ to $2\cdot 10^8\,\mbox{sec.}$ ($1000\,\mbox{years}$).
Thus the measurement of the mass and lifetime of gluino should allow one to estimate
the value of $M_S$ in the Split SUSY scenario.

\subsection{Models with low SUSY breaking scale}
The observed value of the dark energy density can be also reproduced
when the SUSY breaking scale is around $1\,\mbox{TeV}$. This can be
achieved if the MSSM particle content is supplemented by an extra
pair of $5+\bar{5}$ supermultiplets which are fundamental and antifundamental
representations of the supersymmetric $SU(5)$ GUT.
The additional bosons and fermions would not affect gauge coupling unification
in the leading approximation, since they form complete representations of $SU(5)$.
In the physical phase states from $5+\bar{5}$ supermultiplets can gain masses around
$M_S$. The corresponding mass terms in the superpotential can be induced because of
the presence of the bilinear terms $\left[\eta (5\cdot \overline{5})+h.c.\right]$ in
the K$\Ddot{a}$hler potential of the observable sector \cite{30, 31}. In the
Split SUSY scenario we assume that new bosonic states from $5+\bar{5}$ supermultiplets
gain masses around the supersymmetry breaking scale, whereas their
fermion partners acquire masses of the order of the gluino, chargino and neutralino masses.
In our numerical studies we set the masses of extra quarks to be equal to the gluino mass,
i.e. $M_q\simeq M_g$.
In the supersymmetric Minkowski vacuum new bosons and fermions from
$5+\bar{5}$ supermultiplets remain massless.
As a consequence they give a substantial contribution to the $\beta$ functions in
this vacuum. Indeed, the one--loop beta function of the strong interaction in the
second phase changes from $b_3=-3$ (the $SU(3)_C$ beta function in the MSSM) to
$b_3=-2$. This leads to a further reduction of $\Lambda_{SQCD}$.
At the same time, extra fermion states from $5+\bar{5}$ supermultiplets do not
affect much the RG flow of gauge couplings in the physical phase below the scale $M_S$.
For example, in the Split SUSY scenario the one--loop beta function that determines the
running of the strong gauge coupling from the SUSY breaking scale down to the TeV scale
changes from $-5$ to $-13/3$.  As follows from Figs.~\ref{fig1} and \ref{fig2} in
the case of the SUSY model with extra $5+\bar{5}$ supermultiplets the measured value
of the dark energy density can be reproduced even for $M_S\simeq 1\,\mbox{TeV}$
\cite{Froggatt:2005nb, Froggatt:2007qs, Froggatt:2009pj, Froggatt:2010iy, Froggatt:2011fc, Froggatt:2013aua}.
Nevertheless, the Split SUSY scenario which was discussed in the previous subsection has
the advantage of avoiding the need for any new particles beyond those of the MSSM,
provided that $M_S\simeq 10^9 - 10^{10}\,\mbox{GeV}$. On the other hand,
the MPP scenario with extra $5+\bar{5}$ supermultiplets of matter and SUSY
breaking scale in a few $\mbox{TeV}$ range is easier to verify at the LHC
in the near future.

\subsection{The breakdown of SUSY in the hidden sector}
The non-zero value of the vacuum energy density can be also induced if supersymmetry
in the second phase is broken in the hidden sector. This can happen if the SM gauge
couplings are sufficiently small in the supersymmetric Minkowski vacuum and by one way
or another, only vector supermultiplets associated with unbroken non-Abelian gauge
symmetry remain massless in the hidden sector. Then these vector supermultiplets,
that survive to low energies, can give rise to the breakdown of SUSY in the second phase.
Indeed, at the scale $\Lambda_{X}$,  where the gauge interactions that correspond
to the unbroken gauge symmetry in the hidden sector become strong in the second phase,
a gaugino condensate can be formed. This gaugino condensate does not break global
SUSY. Nonetheless if the gauge kinetic function $f_{X}(z_m)$ has a non-trivial dependence
on the hidden sector superfields $z_m$ then the corresponding auxiliary fields $F^{m}$
can acquire non--zero VEVs
\begin{equation}
F^{z_m}\propto \frac{\partial f_X(z_k)}{\partial z_m}\bar{\lambda}_a\lambda_a+...,
\label{47}
\end{equation}
which are set by $<\bar{\lambda}_a\lambda_a>\simeq \Lambda_{X}^3$. Thus
it is only via the effect of a non-renormalisable term that this condensate
causes the breakdown of supersymmetry. Therefore the SUSY breaking scale
in the SUSY Minkowski vacuum is many orders of magnitude lower than $\Lambda_{X}$,
while the scale $\Lambda_{X}$ is expected to be much lower than $M_{Pl}$.
As a result a tiny vacuum energy density is induced
\begin{equation}
\rho_{\Lambda} \sim \frac{\Lambda_{X}^6}{M_{Pl}^2}\, .
\label{48}
\end{equation}

The postulated exact degeneracy of vacua implies then that the physical phase has
the same energy density as the second phase where the breakdown of local SUSY
takes place near $\Lambda_{X}$. From Eq.~(\ref{48}) it follows that in order to reproduce
the measured cosmological constant the scale $\Lambda_{X}$ has to be somewhat close
to $\Lambda_{QCD}$ in the physical vacuum, i.e.
\begin{equation}
\Lambda_{X}\sim \Lambda_{QCD}/10\,.
\label{49}
\end{equation}
Although there is no compelling reason to expect that $\Lambda_{X}$ and $\Lambda_{QCD}$
should be related, one may naively consider $\Lambda_{QCD}$ and $M_{Pl}$ as the two most
natural choices for the scale of dimensional transmutation in the hidden sector.

\begin{figure}[h]
\begin{center}
\includegraphics[width=3.5in]{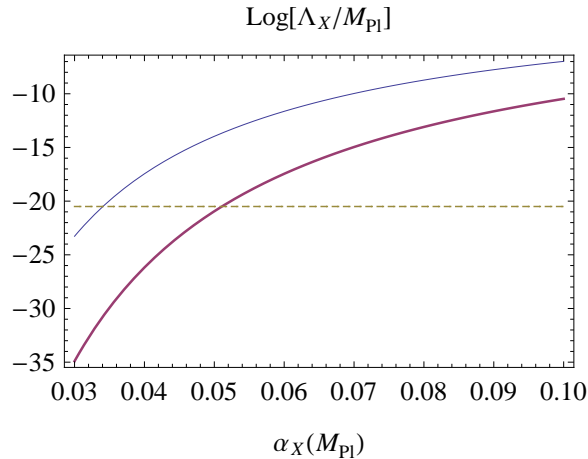}
\end{center}
\caption{The value of $\log\left[\Lambda_{X}/M_{Pl}\right]$ as a function of $\alpha_X(M_{Pl})$.
The thin and thick solid lines are associated with the $SU(3)$ and $SU(2)$ gauge symmetries, respectively.
The horizontal line corresponds to the value of $\Lambda_{X}$ that results in the measured
cosmological constant.}
\label{fig3}
\end{figure}

In the one--loop approximation one can estimate the value of the energy scale $\Lambda_{X}$
using the simple analytical formula
\begin{equation}
\Lambda_{X}=M_{Pl}\exp\left[{\frac{2\pi}{b_X \alpha_X(M_{Pl})}}\right]\,,
\label{50}
\end{equation}
where $ \alpha_X(M_{Pl})=g^2_X(M_Pl)/(4\pi)$, $g_X$ and $b_X$ are the gauge coupling and one--loop beta function
of the gauge interactions associated with the unbroken non-Abelian gauge symmetry that survive to low energies
in the hidden sector. For the $SU(3)$ and $SU(2)$ gauge groups $b_X=-9$ and $-6$, respectively.
In Fig.~\ref{fig3} we show the dependence of $\Lambda_{X}$ on $\alpha_X(M_{Pl})$. As one might expect,
the value of the energy scale $\Lambda_{X}$ diminishes with decreasing $\alpha_X(M_{Pl})$. The observed value of
the dark energy density is reproduced when $\alpha_X(M_{Pl})\simeq 0.051$ in the case of the SUSY model based
on the $SU(2)$ gauge group and $\alpha_X(M_{Pl})\simeq 0.034$ in the case of the $SU(3)$ SUSY gluodynamics.
It is worth noting that in the case of the $SU(3)$ SUSY model the value of the gauge coupling $g_X(M_{Pl})\simeq 0.654$,
that leads to $\alpha_X(M_{Pl})\simeq 0.034$, is just slightly larger than the value of the QCD gauge coupling at the
Planck scale in the SM, i.e. $g_3(M_{Pl})=0.487$\cite{Buttazzo:2013uya}.

In this scenario SUSY can be broken at any scale in the physical vacuum. In particular, the breakdown of local supersymmetry can take
place near the Planck scale. If this is the case, one can explain the small values of $\lambda(M_{Pl})$ and $\beta_{\lambda}(M_{Pl})$
by postulating the existence of a third degenerate vacuum. In this third vacuum local SUSY can be broken near the Planck scale while
the EW symmetry breaking scale can be just a few orders of magnitude lower than $M_{Pl}$. Since now the Higgs VEV is somewhat close to
$M_{Pl}$ one must take into account the interaction of the Higgs and hidden sector fields. Thus the full scalar potential takes the form
\begin{equation}
V=V_{hid}(z_m) + V_0(H) + V_{int}(H, z_m)+...\,,
\label{51}
\end{equation}
where $V_{hid}(z_m)$ is the part of the full scalar potential associated with the hidden sector, $V_0(H)$ is the part of the scalar potential
that depends on the SM Higgs field only and $V_{int}(H, z_m)$ describes the interactions of the SM Higgs doublet with the fields of the hidden sector.
Although in general $V_{int}(H, z_m)$ should not be ignored the interactions between $H$ and hidden sector fields can be quite weak
if the VEV of the Higgs field is substantially smaller than $M_{Pl}$ (say $\langle H \rangle \lesssim M_{Pl}/10$) and the couplings of the
SM Higgs doublet to the hidden sector fields are suppressed.  Then the VEVs of the hidden sector fields in the physical and third vacua can be
almost identical. As a consequence, the gauge couplings and $\lambda(M_{Pl})$ in the first and third phases should be basically the same
and the value of $|m^2|$ in the Higgs effective potential can be still much smaller than $M_{Pl}^2$ and $\langle H^{\dagger} H\rangle$
in the third vacuum. In this limit $V_{hid}(z^{(3)}_m)\ll M_{Pl}^4$ and the requirement of the existence of the third vacuum with vanishingly
small energy density again implies that $\lambda(M_{Pl})$ and $\beta_{\lambda}(M_{Pl})$ are approximately zero in the third vacuum.
Because in this case the couplings in the third and physical phases are basically identical, the presence of such a third vacuum should result
in the predictions (\ref{12}) for $\lambda(M_{Pl})$ and $\beta_{\lambda}(M_{Pl})$ in the physical vacuum.

\section*{Acknowledgments}
This work was supported by the University of Adelaide and the Australian Research Council through the ARC
Center of Excellence in Particle Physics at the Terascale and through grant LF0 99 2247 (AWT). HBN thanks
the Niels Bohr Institute for his emeritus status. CDF thanks Glasgow University and the Niels Bohr Institute
for hospitality and support.

\end{document}